\documentclass[showpacs,twocolumn,amsmath,superscriptaddress]{revtex4-1}
\usepackage[dvipdfmx]{graphicx}
\usepackage{color,amsmath,lmodern,bm,array,braket,booktabs,longtable,here}

\newcolumntype{C}[1]{>{\centering\arraybackslash}p{#1}}
\newcolumntype{L}[1]{>{\raggedright\arraybackslash}p{#1}}
\newcolumntype{R}[1]{>{\raggedleft\arraybackslash}p{#1}}

\begin{document}

\preprint{perspective}

\title{High-throughput calculations of antiferromagnets hosting anomalous transport phenomena}%

\author{Takuya Nomoto}%
\email{nomoto@ap.t.u-tokyo.ac.jp}%
\affiliation{Research Center for Advanced Science and Technology, University of Tokyo, Komaba 153-8904, Japan}
\author{Susumu Minami}%
\affiliation{Department of Mechanical Engineering and Science, Kyoto University, Kyoto 615-8540, Japan}
\author{Yuki Yanagi}%
\affiliation{Liberal Arts and Sciences, Toyama Prefectural University, Toyama 939-0398, Japan}
\author{Michi-To Suzuki}%
\affiliation{Center for Computational Materials Science, Institute for Materials Research, Tohoku University, Sendai 950-8577, Japan}
\affiliation{Center for Spintronics Research Network, Graduate School of Engineering Science,
Osaka University, Toyonaka, Osaka 560-8531, Japan}
\author{Takashi Koretsune}%
\affiliation{Department of Physics, Tohoku University, Sendai 980-8578, Japan}
\author{Ryotaro Arita}%
\affiliation{Research Center for Advanced Science and Technology, University of Tokyo, Komaba 153-8904, Japan}
\affiliation{Center for Emergent Matter Science, RIKEN, Wako 351-0198, Japan}
\date{\today}%

\begin{abstract}
We develop a high-throughput computational scheme based on cluster multipole theory to identify new functional antiferromagnets. This approach is applied to 228 magnetic compounds listed in the AtomWork-Adv database, known for their elevated N\'eel temperatures. We conduct systematic investigations of both stable and metastable magnetic configurations of these materials. Our findings reveal that 34 of these compounds exhibit antiferromagnetic structures with zero propagation vectors and magnetic symmetries identical to conventional ferromagnets, rendering them potentially invaluable for spintronics applications. By cross-referencing our predictions with the existing MAGNDATA database and published literature, we verify the reliability of our findings for 26 out of 28 compounds with partially or fully elucidated magnetic structures in the experiments. These results not only affirm the reliability of our scheme but also point to its potential for broader applicability in the ongoing quest for the discovery of new functional magnets.
\end{abstract}

\maketitle

\section{Introduction}
In computer organization, the importance of memory hierarchy and non-von Neumann architectures has been increasingly recognized, particularly from the perspective of optimizing computational speed and power consumption. This entails the development of various devices optimized for diverse purposes, and innovative material designs are sought after in the field of materials science. Spintronics is a prominent example, which has emerged as a viable option by offering new devices that complement or even replace semiconductor-based devices.

More recently, spintronics utilizing antiferromagnets (AFMs) has gained significant attention since they offer distinct advantages over conventional ferromagnets (FMs). These benefits include the absence of stray field, fast response, and robustness against magnetic field noise~\cite{Jungwirth2016, Baltz2018, Fukami2020, Xiong2022}. However,  the detection and control of magnetic structures that maintain time-reversal symmetry (TRS) present challenges, and thus, particularly desirable for device applications are AFMs breaking TRS. The examples include magnetic toroidal orders~\cite{Watanabe2018, Godinho2018, Vsmejkal2022} and altermagnets~\cite{Naka2019, Vsmejkal2020, Libor2022, Vsmejkal2022_2}, a third class of magnets with zero net magnetization and alternating spin splitting in momentum space. These materials hold promising potential for future applications relying on their controllability via electric currents and the spin splitter effects, respectively.

Among these compounds, AFMs with magnetic symmetries identical to those of conventional FMs are particularly valuable for integration into existing FM-based spintronics devices. These 'FM-like AFMs,' as we term in this paper, feature a small canting moment in the presence of the spin-orbit coupling. Their antiferromagnetic domains can be detected as the sign of anomalous transverse transport and manipulated using magnetic fields. A notable example is Mn$_3X$ ($X=$ Sn and Ge)~\cite{Nakatsuji2015, Nayak2016, Liu2017, Higo2018}, where recent studies have reported magnetization switching via electric current and tunneling magnetoresistance effects~\cite{Tsai2020, Higo2022, Qin2023, Chen2023, Yoon2023}, thereby opening a new era of AFM-based spintronics. However, for practical implementation, numerous considerations such as fabrication procedures, operational temperature ranges, resistance to oxidation, and basic performance metrics intersect in a complex manner. Consequently, the exploration of new functional AFMs continues to be a critical focus for advancing AFM-based spintronics. Indeed, high-throughput calculations for magnetic materials have been vigorously pursued in recent years~\cite{Balluff2017, Horton2019, Zhang2023}. Many of them have focused on specific groups of materials, such as Heusler compounds and two-dimensional materials, and have yet to conduct magnetic symmetry analysis related to the functionality of the AFMs. Moreover, systematic calculations including non-collinear magnetic structures are still challenging, which is required to find prototype FM-like AFM, Mn$_3$Sn.

In this study, we performed high-throughput calculations in exploring potential AFMs for spintronics applications. Note that our present scheme takes into account not only collinear but also non-collinear magnetic structures, which is in clear contrast to the previous studies. Here, we utilized the cluster multipole (CMP) method and spin density functional theory (SDFT) calculations to predict stable magnetic structures from input crystal structures~\cite{Suzuki2017, Suzuki2019, Mita2022, Yanagi2023}. For material pre-screening, we consulted the AtomWork-Adv database~\cite{atomwork}, where AFMs known with specified N\'eel temperature are listed. Subsequent application of the CMP+SDFT methodology on selected candidates allowed us to systematically investigate the most stable and metastable AFM configurations. From the energy comparison and the magnetic space group analysis, we identified 8 metallic and 26 insulating FM-like AFMs as among the most stable magnetic structures. Notably, our findings corroborate the efficacy of our computational approach, as they include several known FM-like AFMs such as Mn$_3$Ir, Mn$_3$Sn, and Mn$_3$SnN. While most identified magnetic structures of FM-like AFMs are already established in the experiments, we discovered that MnPtAl and Fe[NCN] are promising candidates for FM-like AFMs, whose magnetic structures are not completely determined by the experiments. Furthermore, we proposed a range of metastable FM-like AFMs that could potentially be realized through efforts like chemical doping or synthesis optimization. Regarding the effects of electronic correlation, we also performed an analysis using SDFT+$U$ calculations. While we did not discover any new promising FM-like AFM materials, there was considerable overlap between the two sets of calculations. This suggests that the correction from the Hubbard $U$ is not significant, and thus, our scheme including metastable states might be enough to ensure that no potential FM-like AFMs are missed. Our results affirm the effectiveness of our methodology in identifying functional AFMs, even extending to metastable configurations. Future work encompassing a broader spectrum of crystals and magnetic structures with finite propagation vectors could reveal an extensive array of novel functional AFM compounds.

\section{Methods and Workflow}
We performed the CMP+SDFT method to identify functional AFMs~\cite{Mita2022}. The specific workflow adopted in this study is outlined as follows: Initially, we selected input crystal structures for analysis from the AtomWork-Adv database. Secondly, magnetic structures were generated for each crystal using the CMP method to serve as initial inputs for SDFT calculations. These calculations were then executed, and their outputs were subjected to magnetic symmetry analysis for further scrutiny. Below, we briefly summarize each part of the workflow.

\subsection{Input crystal structures}
Although the CMP+SDFT method could accommodate any crystal structure, computational limitation needs a more selective approach to input. For this study, we confined our analysis to structures cataloged in the AtomWork-Adv database~\cite{atomwork}, provided by the National Institute for Materials Science, for several reasons. Firstly, the database contains abundant materials whose N\'eel temperatures, $T_N$, are established, enabling a targeted focus on high-$T_N$ materials, which is crucial for spintronics applications. Secondly, it includes both AFMs with established magnetic structures and those where only a Néel transition has been indicated. This makes the database invaluable both for predicting new AFMs and as a benchmark for our method. 

In practice, we adhere to the following three criteria in selecting crystal structures for analysis: (1) a N\'eel temperature exceeding 273 K, (2) a unit cell containing between 2 and 24 magnetic atoms, and (3) the absence of disorder or partially occupied sites. The condition (1) and (2) are related to the limitation of the computational source as mentioned above. Note that, for the same reason, we consider magnetic structures with zero propagation vector, and thus, the structures with a single magnetic atom in the unit cell do not support AFM states. Since the AFMs with finite propagation vectors are commonly observed in experiments, it is anticipated that in the future, this issue will be addressed combining with the CMP method and the downfolding techniques to derive the classical spin models, such as the local force method~\cite{Liechtenstein1984, Gyorffy1985, Nomoto2020}. The condition (3) is required due to the lack of reliable methods for handling disorder or partial occupancy, such as the coherent potential approximation. While lifting this restriction could be valuable, especially given that chemical substitution is a key strategy for material optimization, we defer this issue to future research. Another point overlooked in this scheme is the impact of finite temperature effects because they cannot be handled in the SDFT calculations. To deal with these effects, we would need to use methods that require more computing cost, such as the disordered local moment method~\cite{Staunton1984}. Alternatively, we can combine our current scheme with a mapping method to the classical spin models. This approach allows us to conduct spin-wave analysis, akin to crystal structure optimization that takes into account the finite temperature effects by including the phonon contributions~\cite{masuki2022}.

\subsection{Input magnetic structures}
In addition to the crystal structures, magnetic structures are required as input for SDFT calculations. For this purpose, we employ the CMP expansion method, as outlined in Refs.~\cite{MTSuzuki2019, YYanagi2023}. This method produces an orthonormal and complete basis set in a 3$N$-dimensional vector space, where $N$ represents the number of magnetic atoms considered for the expansion. This basis set can represent any magnetic structure that arises in a given crystal, as it is classified by multipole degrees of freedom and the irreducible representations of the point group symmetry. Notably, as detailed in Ref.~\cite{Mita2022}, this basis set effectively represents magnetic structures observed in nature. In fact, more than 90\% of magnetic structures cataloged in the MAGNDATA database~\cite{MAGNDATA1,MAGNDATA2} can be expressed by linear combinations of up to three CMP basis elements. Thus, in this study, we focus on magnetic structures that can be represented by linear combinations of up to three CMP basis components. For computational feasibility, we adhere to two conditions for these structures: First, as in Ref.~\cite{Mita2022}, we limit ourselves to linear combinations of CMP bases that belong to the same multipole rank and irreducible representation. Second, we exclude magnetic structures where equivalent atoms (in terms of Wyckoff positions) have different sizes of local moments. Such configurations are often computationally unstable due to the need for spontaneous symmetry breaking in the charge density distribution. While such input magnetic structures can give rise to more plausible AFM structures through the self-consistent iteration, we choose to simplify our model by not considering this possibility.

\subsection{SDFT calculations}
For a given input of crystal and magnetic structure, we perform SDFT calculations to obtain the total energies. We use \texttt{atomate} package for the management of workflow of the high-throughput calculation~\cite{atomate}, which internally utilizes \texttt{pymatgen}, \texttt{custodian}, and \texttt{FireWorks} libraries~\cite{pymatgen, fireworks}. For the electronic structure calculations, we use Vienna {\it ab initio} simulation package (VASP)~\cite{Kresse1996}. Here, we employ the exchange-correlation functional proposed by Perdew, Burke, and Ernzerhof~\cite{Perdew1996}, and pseudopotentials with the projector augmented wave (PAW) basis~\cite{Bloechel1994, Kresse1999}. The spin-orbit coupling effect is also taken into account. The other key setting of the VASP inputs are given as follows: \texttt{ENCUT} = $1.5\times$ (the highest \texttt{ENMAX} in POTCAR file), \texttt{EDIFF} = $10^{-5}$, \texttt{SIGMA} = 0.05, \texttt{LASPH} =  True, \texttt{AMIX} = 0.1, \texttt{AMIX\_MAG} = 0.4, \texttt{BMIX} = 0.0001, \texttt{BMIX\_MAG} = 0.0001, \texttt{PREC} = Accurate, and \texttt{NELM} = 400. We use $\Gamma$ centered $\bm k$-grid generated by \texttt{pymatgen} code with the grid density = $4000/$(number of atoms in the unit cell). For the unconverged cases, we try additional calculations with the change guided by the default setting in \texttt{custodian} code. In the case for SDFT+$U$ calculations, we set $U=3$~eV for all 3$d$ transition metal elements and $0$~eV for the other elements. The other settings are the same as the SDFT calculations.

In this paper, we deal with all transition elements as magnetic and the other elements as non-magnetic. Owing to this definition, we treat the rare-earth elements as non-magnetic although they show magnetic orders in most cases. This is because we consider only materials with high $T_N > 273$ K in this paper, and thus, almost all materials include transition metal elements, as shown in Sec.~\ref{inv_mat}. Even in compounds containing both types of elements, it is a reasonable assumption that only transition metal elements exhibit substantial magnetic moments at elevated temperatures, as commonly seen in rare-earth and transition metal oxides. Accordingly, we employ open-core pseudopotentials for rare-earth elements through the calculations.

\subsection{Analysis of outputs}
Finally, we analyze the total energies and magnetic structures for each material from the SDFT solutions. Here, we regard a material to be AFM when the output magnetic structures satisfy, \begin{align}
\frac{||\sum_i \bm m_i^{\rm out}||}{\sum_i||\bm m_i^{\rm out}||} < 0.05,
\end{align}
where $\bm m_i^{\rm out}$ denotes the magnetic moment of the $i$ site. Then, we identify potential FM-like AFMs by analyzing their magnetic point group symmetries using the \texttt{spinspg} package~\cite{spglib2}, built upon \texttt{spglib}~\cite{spglib1}. Note that a small canting moment is generally induced in the presence of the spin-orbit coupling if thier magnegtic point groups are identical to conventional FMs~\cite{Cheong2024}. However, the anomalous Hall transport is allowed regardless of the existence of net magnetization. In this analysis, we use the input magnetic structures, rather than the outputs, since the self-consistent calculation could break the crystal symmetry due to numerical errors. Instead, we only adopt results where the overlap between the input and output magnetic structures satisfies,
\begin{align}
\bigg|\frac{\sum_i\bm m_i^{\rm in}\cdot \bm m_i^{\rm out}}{\sqrt{\sum_i||\bm m_i^{\rm out}||\sum_i||\bm m_i^{\rm in}||}}\bigg| > 0.99.
\end{align}

Note that not only the most stable but also the metastable magnetic structures are considered in the analysis since they are potentially accessible through experimental techniques such as synthesis optimization, chemical doping, and device-imposed strain. Here, we define the most stable structures as those with a relative energy per magnetic atom, $\Delta E$, less than 0.1 meV, reflecting the numerical accuracy of our high-throughput calculations. Structures with 0.1 meV $<\Delta E<20$ meV are classified as metastable, which approximates the energy scale of the N\'eel temperatures under consideration.

\section{Results and discussion}
\subsection{Investigated materials}\label{inv_mat}
After applying the selection criteria, 231 material candidates remain for consideration. As anticipated, a vast majority of these materials include 3$d$-transition metal elements due to their high N\'eel temperatures. Indeed, only three exceptions, GeS ($T_N=455$K)~\cite{Kyriakos1991}, GdI$_2$ ($T_N=313$K)~\cite{Kasten1984}, and hcp-Gd ($T_N=293$K) are listed in AtomWork-Adv. However, recent literature suggests that GeS may have AFM-like behavior induced by transition metal impurities~\cite{Kyriakos1991}, and GdI$_2$ is ferromagnetic with a Curie temperature $T_c=276$K~\cite{Felser1999, Ahn2000}. Additionally, it is known that hcp-Gd shows ferromagnetism at the highest transition temperature. Thus, these three materials are simply excluded from subsequent calculations. A mere two substances, BaRuO$_3$ ($T_N=430$K) and Sr$_2$MgReO$_6$ ($T_N=320$K), are found to contain only 4$d$ and 5$d$ transition metals as magnetic atoms, respectively. As a result, our following calculations include 226 materials with 3$d$, one with 4$d$, and one with 5$d$ transition metal magnetic materials, which are listed in the Appendix.

In the selected materials, Mn is the most prevalent among 3$d$ transition metals, appearing 100 times. It is followed by Fe with 76 occurrences, Co with 22, Cr with 11, and Ni appearing 10 times. This distribution aligns well with the tendency of Mn and Fe to adopt high-spin configurations in oxide environments, where ions of 3$d$ transition metals typically have valence states of 2+ or 3+. Such configurations often lead to strong exchange interactions and high N\'eel temperatures. This usually results in strong exchange interactions and elevated N\'eel temperatures. It is worth noting that compounds containing Co, as well as Fe, are renowned for their high Curie temperatures, especially in the context of permanent magnets~\cite{Nelson2019}. However, in our analysis of AFMs, Co appears only about one-fourth as frequently as Mn. This is likely due to the AFM nature of 3$d$ transition metals located to the left of Mn, compared to the FM nature of those to the right of Fe~\cite{Sakuma1999}. Therefore, when searching for AFM materials with high Néel temperatures, focusing on Mn-based compounds proves to be the most effective strategy.

\begin{table*}[t] 
     \caption{Summary of FM-like AFMs identified as the most stable magnetic structures found by CMP+SDFT calculations. Substance names  follow the notation of Atomwork-Adv database, where polymorphisms are distinguished by the characters following the formula. $G_{\rm SG}$ represents the space group in standard setting and may not be identical to MAGNDATA/Atomwork-Adv in cartain cases. $G_{\rm MSG}$ represents the magnetic space group of the FM-like AFMs. For systems meeting the criterion $\Delta E<0.1$ meV, all corresponding $G_{\rm MSG}$ symbols are listed. The `Met./Ins.' column signifies whether the system is metallic (M) or insulating (I), based on the calculations. A note is added when a system is metallic in calculations but insulating in experimental data ($^*$M). $T_N$ is the experimental N\'eel temperature, given in Atomwork-Adv. In the `MAGNDATA' column, $\bm q=\bm 0$ and $\bm q\neq\bm 0$ indicate that a magnetic structure with zero and finite propagation vectors is listed in MAGNDATA, respectively. The compounds with $^*1$-$^*4$ indicate that no identical material is listed, but chemically-substituted materials with the same crystal structure are found. Specifically, $^*$1 refers to La$_{0.5}$Sr$_{0.5}$FeO$_{2.5}$F$_{0.5}$, $^*$2 to La$_{0.33}$Sr$_{0.67}$FeO$_3$, $^*$3 to Ho$_{x}$Bi$_{1-x}$FeO$_3$ ($x=0.15, 0.20$), and $^*$4 to Ho$_{x}$Bi$_{1-x}$FeO$_3$ ($x=0.05, 0.10, 0.15, 0.20$).  The 'conv. AFM' column specifies whether there is a conventional AFM structure also meets the criterion $\Delta E<0.1$ meV. An asterisk $(^*)$ indicates that a conventional AFM is indicated in the experiments. The `Consistency' column employs two symbols to denote the following: $\times$ represent that a $\bm q=0$ structure is reported in the experiments, but our method fails to reproduce it as the most stable states; an asterisk ($*$) indicates that a $\bm q\neq0$ structure is reported in the experiments, which is beyond the reach of our method. All other entries imply that the magnetic structures have either been partially or fully corroborated by the experiments and are also reproduced by our method.} 
     \label{tab:fm-afm1}
    \centering
    \begin{tabular}{wc{2.5cm}wc{2cm}wc{2.5cm}wc{2cm}wc{1.5cm}wc{2cm}wc{2cm}wc{2cm}} 
        \toprule[.3mm]
        Substance & $G_{\rm SG}$ & $G_{\rm MSG}$& Met./Ins. & $T_{\rm N}$ [K] & MAGNDATA & conv. AFM & Consistency\\
        \midrule[.15mm]
        Mn$_{3}$Ir rt & $Pm\bar{3}m$ & $R\bar{3}m'$ & M & 855 & $\bm q=\bm 0$ & No & \\
        LaFeO$_{3}$ rt & $Pnma$ & $Pn'ma', Pn'm'a$ & I & 746 & $^{*1}\bm q=\bm 0$ & Yes  & \\
        LaFeO$_{3}$ ht$_{1}$ & $R\bar{3}c$ & $P\bar{1}, C2/c$ & I & 740 & $^{*2}\bm q\neq\bm 0$ & Yes & $*$ \\
        PrFeO$_{3}$ orth & $Pnma$ & $Pn'ma', Pn'm'a$ & I & 730 & - & Yes &  \\
        Ca$_{2}$Fe$_{2}$O$_{5}$ rt & $Pnma$ & $Pn'm'a$ & I & 723 & $\bm q=\bm 0$ & No & \\
        EuFeO$_{3}$ & $Pnma$ & $Pn'ma', Pn'm'a$ & I & 689 & - & Yes &  \\
        SmFeO$_{3}$ & $Pnma$ & $Pn'ma', Pn'm'a$ & I & 688 & $\bm q=\bm 0$ & Yes &  \\
        Sr$_{2}$Fe$_{2}$O$_{5}$ rt & $Ima2$ & $Im'a2'$ & I & 680 & - & No & $\times$ \\
        KFe[MoO$_{4}$]$_2$ hp$_{3}$ & $P\bar{3}c1$ & $C2'/c', C2/c$ & I & 653 & - & No & $*$\\
        FeBiO$_{3}$ hp V & $Pnma$ & $Pn'ma'$ & I & 643 & $^{*3}\bm q=\bm 0$ & No &  \\
        YFeO$_{3}$ rt & $Pnma$ & $Pn'ma', Pn'm'a$ & I & 639 & - & Yes &  \\
        HoFeO$_{3}$ & $Pnma$ & $Pn'ma', Pn'm'a$ & $^*$M & 637 & $^{*3}\bm q=\bm 0$ & Yes & \\
        ErFeO$_{3}$ & $Pnma$ & $Pn'ma', Pn'm'a$ & I & 636 & - & Yes &  \\
        NaFe[MoO$_{4}$]$_2$ & $C2/c$ & $C2/c$ & I & 635 & - & No & $*$\\
        TmFeO$_{3}$ orth & $Pnma$ & $Pn'ma', Pn'm'a$ & I & 630 & - & Yes &  \\
        YbFeO$_{3}$ orth & $Pnma$ & $Pn'ma', Pn'm'a$ & I & 629 & - & Yes & \\
        FeBiO$_{3}$ rt & $R3c$ & $P1, Cc$ & I & 627 & $^{*4}\bm q=\bm 0$ & No  &  \\
        NdFeO$_{3}$ & $Pnma$ & $Pn'ma', Pn'm'a$ & I & 609 & $\bm q=\bm 0$ & Yes &  \\
        FeS ht$_{2}$ & $P6_3/mmc$ & $Cm'c'm$ & $^*$M & 599 & - & Yes &  \\
        TlFeO$_{3}$ & $Pnma$ & $Pn'ma', Pn'm'a$ & $^*$M & 560 & - & Yes & \\
        DyFeO$_{3}$ & $Pnma$ & $Pn'ma', Pn'm'a$ & I & 540 & $\bm q=\bm 0$ & Yes &  \\
        GdFeO$_{3}$ & $Pnma$ & $Pn'ma', Pn'm'a$ & I & 530 & - & Yes & \\
        Fe$_{3}$BO$_{6}$ & $Pnma$ & $Pnm'a'$ & I & 508 & - & No & \\
        La$_{2}$NiO$_{4}$ lt & $P4_2/ncm$ & $P4_2/nc'm'$ & I & 420 & $\bm q=\bm 0$ & Yes & $\times$ \\
        Mn$_{3}$Sn & $P6_3/mmc$ & $Cmc'm', Cm'cm'$ & M & 411 & $\bm q=\bm 0$ & No & \\
        Mn$_{3}$SnN & $Pm\bar{3}m$ & $R\bar{3}m'$ & M & 395 & - & No  &  \\
        FeF$_{3}$ rt & $R\bar{3}c$ & $P\bar{1}, C2/c$ & I & 364 & $\bm q=\bm 0$ & $^*$Yes & \\
        Fe[NCN] & $P6_3/mmc$ & $Cm'c'm$ & I & 345 & - & Yes & \\
        LaCrO$_{3}$ ht$_{1}$ & $R\bar{3}c$ & $C2/c$ & I & 327 & $\bm q=\bm 0$ & $^*$No & \\
        Nd$_{2}$NiO$_{4}$ rt & $Cmce$ & $Cm'ca'$ & I & 323 & $\bm q\neq\bm 0$ & Yes & $*$ \\
        LaCrO$_{3}$ rt & $Pnma$ & $Pn'ma', Pn'm'a$ & I & 291 & $\bm q=\bm 0$ & Yes & \\
        MnPtAl & $P6_3/mmc$ & $Cm'c'm$ & M & 290 & - & No & \\
        AgMn$_{3}$N & $Pm\bar{3}m$ & $R\bar{3}m'$ & M & 281 & - & $^*$Yes & \\
        SrMnO$_{3}$ 4H rt & $C222_1$ & $C22'2_1'$ & I & 280 & - & Yse &  \\
        \bottomrule[.15mm]
    \end{tabular}
\end{table*}

\subsection{Crystal structure analysis}
In our dataset of 228 materials, several share identical crystal structure prototypes. To understand the structural types more likely to exhibit high N\'eel temperatures, we performed a prototype analysis. In this work, two conditions define an equivalence relation for classifying these prototypes: (1) the space group number and (2) the set of atom counts in the unit cell, although exact matching at the Wyckoff positions is not required. 

The results are summarized as follows. The most frequently occurring crystal prototype is the ThCr$_2$Si$_2$ structure with space group $G_{\rm SG}=$ $I4/mmm$, appearing 35 times. Most of them are $R$Mn$_2$Si$_2$ and $R$Mn$_2$Ge$_2$ ($R$ is rare-earth elements), and unfortunately, FM-like AFM structures with zero propagation vector are hardly realized in this prototype. The second most common is the HfFe$_6$Ge$_6$ structure with $G_{\rm SG}=$ $P6/mmm$, appearing 27 times. This structure has been the subject of intense recent study due to its unique properties related to the Kagome net of the Fe layer. The third most common is the orthorhombic perovskite $AB$O$_3$, appearing 18 times, followed by the CeCo$_4$B, tetragonal CeFeSi, and HfFe$_6$Sn$_4$Ge$_2$ structures, each appearing 8 times. The orthorhombic TiNiSi appears 7 times, and the NiAs structure 6 times. Among these frequently appearing prototypes, only the HfFe$_6$Ge$_6$ and HfFe$_6$Sn$_4$Ge$_2$ structures can host non-collinear AFMs based on our CMP+SDFT analysis. This suggests that high-$T_N$, zero-propagation vector, and non-collinear FM-like AFMs are rarely realized in nature. Therefore, collinear AFMs breaking TRS, such as altermagnets, would play a significant role in future spintronics applications. In the following, we will see that the NiAs structure and its derivatives are such promising crystal structures in the search for FM-like AFMs. 

\subsection{FM-like AFMs as the most stable structures}
Table~\ref{tab:fm-afm1} shows the FM-like AFMs identified as the most stable magnetic structures in the CMP+SDFT method. The 34 materials are listed, where 8 are metals and 26 are insulators, although SDFT calculations sometimes fail to reproduce metallicity due to the approximate treatment of the electronic correlation effects. For 16 of these 34 AFMs, the magnetic structures are already established and listed in the MAGNDATA database, as indicated in Table~\ref{tab:fm-afm1}. Only two compounds, LaFeO$_3$~ht$_1$ and Nd$_2$NiO$_3$~rt, exhibit AFM structures with finite propagation vectors, which are beyond the scope of our current study. However, for 13 out of the remaining 14 compounds, MAGNDATA features at least one magnetic structure that aligns with one of the most stable AFMs determined in our calculations. The sole exception is La$_2$NiO$_4$~lt, where a collinear AFM structure with the magnetic moments aligned parallel to the $a$-axis. In our CMP+SDFT calculations, the closest initial guess has finite components along both the $a$-axis and $c$-axis. After self-consistent calculations, this leads to an AFM structure oriented along the $c$-axis. Since this structure is different from the one listed in MAGNDATA only in the direction of the magnetic moments, it could potentially be realized by some efforts including a chemical doping. It should be noted that although FeF$_3$~rt and LaCrO$_3$~ht$_1$ are identified as conventional AFMs in MAGNDATA, these are also reproduced in our calculations as nearly degenerate states.

Even in cases where magnetic structures are not listed in the MAGNDATA database, there are experimental studies providing direct and indirect evidence for their magnetic structures in most cases. After examining the existing literature, it appears that the magnetic structures of Fe[NCN] and MnPtAl are not fully established in the experiments, in which our results offer predictive insights into the potential FM-like AFM structures. A brief discussion of these compounds is presented below.

\subsubsection{orthorhombic RFeO$_3$}
The orthorhombic perovskites $R$FeO$_3$, where $R=$ Pr, Eu, Y, Er, Tm, Yb, and Gd, are identified as potential FM-like AFMs yet remain unlisted in MAGNDATA. However, these materials are well-studied, particularly for their ferroelectric and multiferroic properties. They generally exhibit two types of magnetic transitions. The first type initially forms a $G_z$-type AFM structure with $G_{\rm MSG}=Pn'ma'$, followed by a spin-reorientation transition to a $G_y$-type AFM with $G_{\rm MSG}=Pn'm'a$. Compounds in this category include $R=$ Pr~\cite{Sosnowska1982,Sosnowska1985,Li2019}, Er~\cite{Gorodetsky1973,Bazaliy2004,Deng2015}, Tm~\cite{Leake1968,kimel2004}, and Yb~\cite{Bazaliy2005,Nikitin2018,Xiaoxuan2021}. The second type undergoes a single magnetic transition to the $G_z$-type AFM, as in the cases of $R=$ Y~\cite{Hahn2014,Rodriguez2019}, Eu~\cite{White1969,Seifu2006}, and Gd~\cite{White1969}. This fact shows that the two AFM structures are energetically so close that they can interchange with a small energy scale of the temperature effects, and our SDFT calculations correctly capture this feature. Note that although these AFM states often possess a weak FM component due to the Dzyaloshinskii-Moriya interaction, which is not accounted for in our initial structures, it appears during the self-consistent calculation. As these compounds are insulators, magneto-optical effects are anticipated for their AFM structures, and these have already been observed experimentally~\cite{Treves1965, Kahn1969}. Note that TlFeO$_3$ is also indicated to be an FM-like AFM in our calculations. While the specifics of its magnetic structure are likely not well-documented, its crystal structure and systematic variations in $T_N$ lead us to expect properties similar to those of $R$FeO$_3$~\cite{Kim2001}.

\subsubsection{Sr$_2$Fe$_2$O$_5$}
The compound Sr$_2$Fe$_2$O$_5$ with space group $G_{\rm SG}=Ima2$ is a brownmillerite oxide. Neutron diffraction experiments suggested that its magnetic structure is consistent with $G_{\rm MSG}=Im'a'2$~\cite{Takeda1969, Schmidt2001}, however, this result is in contradiction with our computational prediction of the FM-like AFM with $G_{\rm MSG}=Im'a2'$. The discrepancy arises because our initial guess does not include the experimentally suggested $G$-type AFM structure. The complexity lies in the fact that such a structure demands a linear combination of non-equivalent Fe moments, which belong to different irreducible representations of the point group. To capture such AFM configurations, it will be necessary to expand the search area by relaxing the present constraints. 

\subsubsection{AFe[MoO$_4$]$_2$}
$A$Fe[MoO$_4$]$_2$, where $A=$ K and Na, are layered molybdates known for their ferroelastic transition from a high-temperature trigonal phase with $G_{\rm SG}=P\bar{3}c1$ to a low-temperature monoclinic phase with $G_{\rm SG}=C2/c$~\cite{Otko1978}. The frustrated triangular Fe net has prompted interest in their magnetic properties, and indeed, the multiferroic AFM structures along with electronic polarization for $R=$ Rb have been intensively discussed recently~\cite{Kenzelmann2007, Mitamura2014}. For the $R=$ K case, coexistence of spin spiral and commensurate AFM structures has been reported, both with finite propagation vectors~\cite{Smirnov2010}, although $T_N$ reported in Ref.~\cite{Smirnov2010} is $2.5$ K considerably lower than $653$ K in Ref.~\cite{Ismailzade1981} referred in AtomWork-Adv. In the Na variant, experimental data on magnetic structures seems to be lacking, but a theoretical investigation suggested a $G$-type AFM with finite propagation vectors as the most stable state~\cite{Tamboli2021}, which our current calculations cannot capture.

\subsubsection{FeS ht$_2$}
FeS is characterized as a $p$-type narrow-gap semiconductor and undergoes a structural transition from a high-temperature hexagonal phase (FeS ht$2$) with $G_{\rm SG}=P6_3/mmc$ to a low-temperature trigonal phase with $G_{\rm SG}=P\bar{6}2c$~\cite{Putnis1974}. Although AtomWork-Adv lists an additional phase, FeS ht$1$ with $G_{\rm SG}=P6_3mc$, we have chosen to exclude it from consideration in this study since it includes the distortion accompanied by the N\'eel transition~\cite{FLi1996}. In both high- and low-temperature phases, FeS manifests collinear AFM order. Within each Fe layer, the local moments align ferromagnetically, while they stack antiferromagnetically along the $c$-axis. Additionally, a spin reorientation transition, commonly referred to as the Morin transition, occurs in the low-temperature phase~\cite{Coey1976, Horwood1976}. In this transition, the local moments reorient from being normal to parallel with respect to the $c$-axis. The easy-plane AFM structure aligns with the calculated $Cm'c'm$ phase. This phase holds the potential for exhibiting anomalous transport phenomena, a claim recently substantiated through experimental work~\cite{Seki1}.

\subsubsection{Fe$_3$BO$_6$}
Fe$_3$BO$_6$ is an insulator that exhibits a collinear AFM structure with a minor FM component. Similar to orthoferrites, $R$FeO$_3$, this compound undergoes a spin reorientation transition between two distinct AFM configurations~\cite{Wolfe1969, Hirano1974}. In the high-temperature phase, the magnetic moments align antiferromagnetically along the $c$-axis, while two non-equivalent Fe moments within the same $c$-plane couple ferromagnetically~\cite{Nakamura2017}. This magnetic structure matches the $G_{\rm MSG}=Pnm'a'$ configuration listed in Table~\ref{tab:fm-afm1}. It is anticipated to exhibit anomalous magneto-optical effects, which is in alignment with experimental findings~\cite{Abe1980}.

\begin{table*}[t]
    \centering
    \caption{List of FM-like AFMs identified as metastable magnetic structures found by CMP+SDFT calculations. $\Delta E$ is the relative energy to the ground state per one magnetic atom of the FM-like AFM structure. The other notations carry the same meaning as in the Table~\ref{tab:fm-afm1}.} \label{tab:fm-afm2}
    \begin{tabular}{wc{3cm}wc{2cm}wc{2.5cm}wc{2.5cm}wc{1.5cm}wc{2cm}wc{2cm}} 
        \toprule[.3mm]
        Substance & $G_{\rm SG}$ & $G_{\rm MSG}$ & $\Delta E$ [meV] & Met./Ins. & $T_{\rm N}$ [K] & MAGNDATA  \\
        \midrule[.15mm]
        Fe$_{2}$O$_{3}$ hem & $R\bar{3}c$ & $P\bar{1}, C2'/c', C2/c$ & 0.14, 0.14, 0.15 & I & 957 & $\bm q=\bm 0$ \\
        CrSb & $P6_3/mmc$ & $Cm'c'm$ & 0.11 & M & 711 & $\bm q=\bm 0$ \\
        FeSe hp$_{1}$ & $P6_3/mmc$ & $Cm'c'm$ & 16.8 & M & 460 & -  \\
        Ca$_{2}$MnFeO$_{5}$ orth$_{1}$ & $Pnma$ & $Pnm'a', Pn'm'a$ & 0.4, 0.3 & I & 407 & -  \\
        Mn$_{3}$SnC & $Pm\bar{3}m$ & $R\bar{3}m'$ & 0.3 & M & 348 & - \\
        MnTe rt & $P6_3/mmc$ & $Cm'c'm$ & 0.2 & I & 309 & $\bm q=\bm 0$ \\
        CrSb$_{2}$ & $Pnnm$ & $Pnn'm'$ & 4.4 & $^*$M & 275 & - \\
        \bottomrule[.15mm]
    \end{tabular}
\end{table*}

\subsubsection{Mn$_3$SnN and AgMn$_3$N}
Mn$_3$SnN and AgMn$_3$N belong to a family of cubic anti-perovskite nitrides. This family typically features two types of magnetic structures within the Kagome (111) plane: two 120-degree triangular spin textures that are rotated by 90 degrees relative to each other~\cite{Fruchart1978}. These configurations are referred to as the $\Gamma_{4g}$ and $\Gamma_{5g}$ structures, whose magnetic symmetries are $G_{\rm MSG}=R\bar{3}m'$ and $R3m$, respectively. Although the energy levels of these two states are close due to a result of the six-fold symmetry of magnetic anisotropy and the small spin-orbit couplings, only the $\Gamma_{4g}$ state exhibits the anomalous Hall effect attributed to nonzero Berry curvature, which has been experimentally observed in Mn$_3$SnN thin films~\cite{Gurung2019,You2020}. More precisely, AgMn$_3$N initially exhibits the $\Gamma_{5g}$ structure before transitioning to a mixed phase of $\Gamma_{4g}+\Gamma_{5g}$ states. Conversely, Mn$_3$SnN starts with the mixed state and later transitions to a more complex phase characterized by finite propagation vectors~\cite{Fruchart1978}. Note that
our calculations focus solely on states with zero propagation vector and exclude mixed states with different irreducible representations, and thus, an exact match with experimental observations cannot be achieved. Nevertheless, both the $\Gamma_{4g}$ and $\Gamma_{5g}$ states are identified as one of the ground state magnetic structures in AgMn$_3$N, and $\Gamma_{4g}$ as the ground state and $\Gamma_{5g}$ as the meta-stable state in Mn$_3$SnN. Although accurately estimating the stability of these structures is challenging due to minor differences in energy scales, these results indicate that our scheme reliably finds potentially useful magnetic structures. Note that the magnetic structures and anomalous Hall effects of this Mn$_3X$N group have been the subject of intensive theoretical investigation in recent years~\cite{Huyen2019,Zhou2019}.

\begin{figure}[t]
\centering
\includegraphics[width=0.45\textwidth]{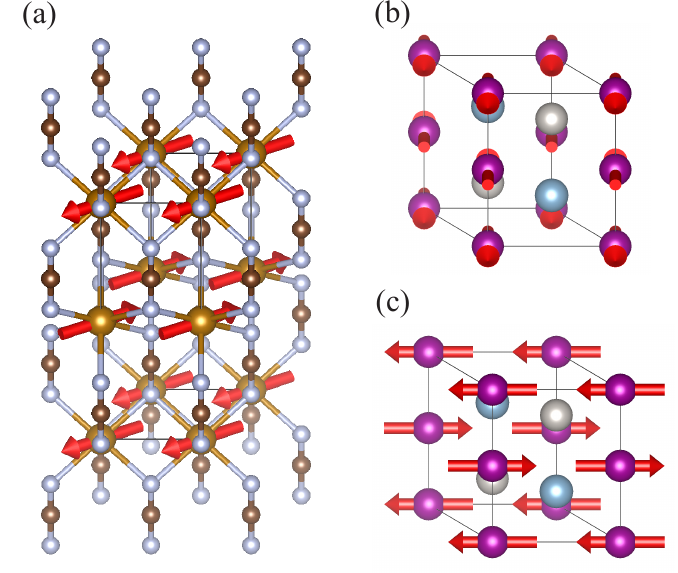}
\caption{
Magnetic configurations of (a) $G_{\rm MSG}=Cm'c'm$ structure of Fe[NCN], (b) $G_{\rm MSG}=Cm'c'm$ structure, and (c) $G_{\rm MSG}=Cmcm$ structure of MnPtAl. 
}
\label{fig:1}
\end{figure}

\subsubsection{Fe[NCN]}
Fe[NCN] is recognized as a potent electrode material in Li- and Na-ion batteries~\cite{Chen2020}. It crystallizes in a hexagonal lattice with space group $G_{\rm SG}=P6_3/mmc$ at room temperature, featuring Fe$^{2+}$ planes interconnected by linear NCN$^{2-}$ chains. While antiferromagnetic coupling of Fe atoms has been suggested in prior studies~\cite{Liu2009, Herlitschke2014}, the exact magnetic structure seems to remain undetermined. Our results, whose magnetic structure is shown in Fig.~\ref{fig:1}(a), indicate that exploring the magneto-optical properties of this material could be quite interesting.

\subsubsection{MnPtAl}
MnPtAl is an intermetallic compound with a Ni$_2$In-type structure, sharing the same crystal arrangement as MnPtGa, known to host a Néel-type skyrmion phase~\cite{Srivastava2020}. Neutron studies suggested that the Mn moments in MnPtAl align ferromagnetically within the plane and antiferromagnetically along the $c$-axis~\cite{Arons1996}. These magnetic moments are perpendicular to the $c$-axis, which is in agreement with the $G_{\rm MSG}=Cm'c'm$ phase predicted in our calculations, shown in Fig.~\ref{fig:1}(b). It is worth noting that, similar to FeS with the NiAs structure, the in-plane direction of the magnetic moments is critically important; a 30-degree rotation of each spin can transform the FM-like AFM into a conventional AFM, shown in Fig.~\ref{fig:1}(c). Thus, the experimental determination of the local moment directions, as well as the observation of anomalous Hall effects, is important, although our calculations indicated that the in-plane magnetic anisotropy is not very weak as no conventional AFM states are found as the nearly degenerate states~\cite{Seki2}.
 
\subsubsection{SrMnO$_3$ 4H rt}
 SrMnO$3$ 4H rt is a distorted hexagonal perovskite with $G_{\rm SG}=C222_1$, and is found to have a magnetic structure identical to the $G_{\rm MSG}=C22'2'_1$ in our calculations. This is consistent with neutron diffraction studies that describe the AFM ordering along the $c$-axis both within the Mn$_2$O$_9$ dimers and between the corner-sharing MnO$_6$ octahedra, with spins parallel to the $a$-axis~\cite{Rune2006,Daoud2007}. Note that a 90-degree rotated AFM structure also falls under the category of FM-like AFM unlike FeS, Fe[NCN], and MnPtAl structures.

\begin{table*}[t]
    \centering
    \caption{Summary of FM-like AFMs identified as the most stable magnetic structures newly found by CMP+SDFT+$U$ calculations. The notations carry the same meaning as in the Table~\ref{tab:fm-afm1}. Note that CrSb and MnTe rt are listed in TABLE II.}\label{fm-afm3}
    \begin{tabular}{wc{2.5cm}wc{2cm}wc{2.5cm}wc{2cm}wc{1.5cm}wc{2cm}wc{2cm}wc{2cm}} 
        \toprule[.3mm]
        Substance & $G_{\rm SG}$ & $G_{\rm MSG}$ & Met./Ins. & $T_{\rm N}$ [K] & MAGNDATA & conv. AFM & consistency \\
        \midrule[.15mm]
        CrSb & $P6_3/mmc$ & $Cm'c'm$ & $^*$M & 711 & $\bm q=\bm 0$ & No \\
        Mn$_{2}$As & $P4/nmm$ & $P4/nm'm'$ & M & 582 & $\bm q\neq\bm 0$ & No & $*$ \\
        Sr$_{2}$Co$_{2}$O$_{5}$ & $Ima2$ & $Ima'2'$ & M & 535 & $\bm q\neq\bm 0$ & No & $*$ \\
        Fe$_{2}$As & $P4/nmm$ & $Pmm'n', Cmm'a'$ & M & 353 & $\bm q\neq\bm 0$ & No & $*$\\
        FeGe rt & $P2_13$ & $P2_1'2_1'2_1$ & M & 310 & $\bm q\neq\bm 0$ & Yes & $*$\\
        MnTe rt & $P6_3/mmc$ & $Cm'c'm$ & I & 309 & $\bm q=\bm 0$ & Yes \\
        NiS ht & $P6_3/mmc$ & $Cm'c'm$ & I & 304 & - & $^*$Yes \\
        \bottomrule[.15mm]
    \end{tabular}
    \vspace{0.5mm}
    \caption{List of FM-like AFMs identified as metastable magnetic structures newly found by CMP+SDFT+$U$ calculations. The notations carry the same meaning as in the Table~\ref{tab:fm-afm2}. Note that FeS ht$_{2}$, Fe[NCN], MnPtAl, and AgMn$_{3}$N are listed in TABLE I.}\label{fm-afm4}
    \begin{tabular}{wc{3cm}wc{2cm}wc{2.5cm}wc{2.5cm}wc{1.5cm}wc{2cm}wc{2cm}}
        \toprule[.3mm]
        Substance & $G_{\rm SG}$ & $G_{\rm MSG}$ & $\Delta E$ [meV] & Met./Ins. & $T_{\rm N}$ [K] & MAGNDATA \\
        \midrule[.15mm]
        FeS ht$_{2}$ & $P6_3/mmc$ & $Cm'c'm$ & 0.9 & I & 599 & -  \\
        Fe[NCN] & $P6_3/mmc$ & $Cm'c'm$ & 2.4 & I & 345 & -  \\
        MnNiGe rt & $Pnma$ & $Pnm'a'$ & 1.3 & M & 336 & - \\
        Ba$_{2}$Y$_{2}$Co$_{4}$O$_{11}$ rt & $Pmmm$ & $Pm'm'm$ & 14.5 & M & 291 & - \\
        MnPtAl & $P6_3/mmc$ & $Cm'c'm$ & 2.5 & M & 290 & - \\
        AgMn$_{3}$N & $Pm\bar{3}m$ & $R\bar{3}m'$ & 0.1 & M & 281 & - \\
        \bottomrule[.15mm]
    \end{tabular}
\end{table*}

\subsection{FM-like AFMs as the metastable structures}
An advantage of the CMP method lies in its ability to systematically assess the magnetic structures with energies of metastable states. Table~\ref{tab:fm-afm2} outlines FM-like AFM materials that we identified as metastable states; these are characterized by $\Delta E$ is less than 20 meV. Table~\ref{tab:fm-afm2} shows 7 such metastable magnetic structures, with experimental data for three of these materials available in MAGNDATA. Thanks to their small $\Delta E$, we expect that it is possible to realize them given the experimental tunability, finite temperature effects, and the limitation of the accuracy of SDFT. For instance, Fe$_2$O$_3$~hem exhibits a collinear AFM and undergoes a Morin transition, altering the direction of magnetic moments from in-plane at high temperatures to out-of-plane at low temperatures~\cite{Hill2008}. These transitions respectively align well with the $C2'/c'$ and $P\bar{1}$ states in Table~\ref{tab:fm-afm2}. MnTe~rt, which adopts the NiAs structure, serves as another such example. This compound has been reported to exhibit an antiferromagnetic (AFM) state with magnetic moments aligned parallel to the $[1\bar{1}00]$ direction, representing a conventional AFM structure~\cite{Kunitomi1964}. However, recent research suggests that strain induced by epitaxial lattice mismatch with a substrate alters the in-plane magnetic anisotropy. Specifically, this leads the magnetic moments to align parallel to the $[11\bar{2}0]$ direction, corresponding to the $Cm'c'm$ state listed in Table~\ref{tab:fm-afm2}~\cite{Kriegner2017}. In fact, this structure has been the subject of recent study as a typical example of an altermagnet~\cite{Betancourt2023}. These observations demonstrate the effectiveness of our approach in identifying metastable structures.

For the other candidates, we confirmed that the experimental structures do not correspond to FM-like AFMs listed in Table~\ref{tab:fm-afm2}. Here, we briefly comment on them. CrSb belongs to the NiAs structure and shows staggered AFM along the $c$-axis, whose magnetic moments are aligned parallel to the $c$-axis~\cite{Snow1953, Yuan2020}. $Cm'c'm$ structure in Table~\ref{tab:fm-afm2} is obtained by rotating spins of it to the in-plane. MnSn$_3$C is known to show complex spin configurations consisting of AFM and AM components with finite propagation vectors, as suggested in Ref.~\cite{Lorthioir1973}. Our results show that $R\bar{3}m'$ structure, which is the same as MnSn$_3$N, also lies as a low energy state. Ca$_2$FeMnO$_5$~orth$_1$ is known to show $G$-type AFM structure, whose magnetic moment is parallel to the $c$-axis~\cite{Rykov2008}. Our $Pnm'a'$ and $Pn'm'a$ structures correspond to the same AFM structure with the moments perpendicular to the $c$-axis like Ca$_2$Fe$_2$O$_5$~\cite{Auckett2015}. CrSb$_2$ is known to show collinear AFM structure with a finite propagation vector~\cite{Holseth1970}, which cannot be accessible in our current scheme. FeSe hp$_1$ is known to show both AFM and FM-like behavior, depending on temperature and amount of self-intercalation of Se~\cite{Song2013}. Since the origin of the AFM nature of FeSb has not been clear yet, we may have a chance to realize an FM-like AFM state in hexagonal FeSe. 

\subsection{FM-like AFMs in SDFT+U calculations}
Finally, we comment on the results with the SDFT+$U$ calculations. As widely recognized, strong  electronic correlation effects play a critical role in many magnetic materials, especially in oxides, and these effects are not sufficiently captured by standard SDFT. Therefore, we explored the impact of the correlation effects on magnetic structures by performing the SDFT+$U$ calculations. According to the systematic calculations, we found that while the inclusion of the Hubbard $U$ does affect the energy levels of various magnetic states, it does not significantly change their order in most materials. Indeed, we identified only 7 FM-like AFMs as ground states that are not found in the SDFT calculations, as shown in TABLE~\ref{fm-afm3}. Of these, 2 are also identified as metastable states in the SDFT calculations. Among the remaining 5 materials, 4 have established AFM structures listed in MANGDATA, all of which have finite propagation vectors, and thus, fall outside the scope of our current scheme. The last one, NiS~ht with the NiAs-structure, is known in experiments to be an AFM oriented along the $c$-axis~\cite{Sparks1963}. This state is indeed found in our calculations as one of the most stable magnetic structures; however, it is a conventional AFM.

The results for metastable states are shown in Table~\ref{fm-afm4}. Our claculations identify 6 FM-like AFMs not identified as metastable states in the SDFT calculations, while 4 of them are already listed as the most stable magnetic structures in the SDFT calculations. The remaining two are MnNiGe rt and Ba$_2$Y$_2$Co$_4$O$_{11}$ rt. MnNiGe shares the same crystal structures as NbMnP, in which the anomalous Hall effect in an AFM state has been recently observed~\cite{Kotegawa2023}. However, MnNiGe is known to be an AFM with a finite propagation vector, making it outside what our calculations can handle~\cite{Fjellvag1985}. Similarly, Ba$_2$Y$_2$Co$_4$O$_{11}$, a layered perovskite, is also known to be an AFM with a finite propagation vector, again exceeding the bounds of our current approach~\cite{Aurelio2007}. In summary, while certain materials have been newly recognized through the SDFT+$U$ calculations, a majority coincide with those identified in the SDFT calculations, suggesting that the incorporation of $U$ exerts small influence on the stability of magnetic structures, particularly when metastable states are also considered in the evaluation.

\section{Conclusion}
In this work, we have introduced a high-throughput computational framework utilizing CMP theory and SDFT calculations to systematically identify new functional AFMs, suitable for spintronics applications. We applied this approach to 228 magnetic compounds cataloged in the AtomWork-Adv database, renowned for their elevated Néel temperatures. Our systematic investigations yielded 34 compounds that display AFM structures with zero propagation vectors, featuring magnetic symmetries identical to conventional FMs. Among these, 9 are metallic and 25 are insulating, and our results include known FM-like AFMs such as Mn$_3$Sn and Mn$_3$SnN. Furthermore, we identified MnPtAl and Fe[NCN] as promising, yet experimentally undetermined, candidates for FM-like AFM materials. By cross-referencing our theoretical predictions with existing experimental data from the MAGNDATA database and published literature, we confirmed the accuracy of our findings for 26 out of the 28 compounds that have been partially or fully characterized in experiments. This validation not only attests to the reliability of our computational scheme but also hints at its broader applicability in the continuous endeavor to discover new functional magnetic materials. Importantly, our method has also uncovered a range of metastable FM-like AFM structures that could be experimentally realized through techniques such as chemical doping or synthesis optimization.

In summary, our study establishes a solid foundation for future research that could expand the scope to include a wider variety of crystal structures and magnetic arrangements with finite propagation vectors, thereby opening new avenues for the discovery of novel functional AFM materials

\section{Acknowledgments}
We are grateful to Dr. Tanno for the technical supports. We also acknowledge the use of supercomputing system, MASAMUNE-IMR, at CCMS, Tohoku University in Japan. This work was financially supported by JSPS-KAKENHI (No. JP23H04869, JP23H01130, JP22K03447, JP22H00290, JP22K14587, JP21H04437, JP21H04990, JP21H01789, JP21H01031, and JP19H05825), JST-Mirai Program (JPMJMI20A1), JST-PRESTO (No. JPMJPR20L7), JST-CREST (No. JPMJCR18T3, JPMJCR23O3), and JST-ASPIRE (No. JPMJAP2317).

\section*{Appendix: List of all materials}
This appendix shows the summary of materials we considered in this paper. These are listed in the AtomWork-Adv database and satisfy (1) a N\'eel temperature exceeding 273 K, (2) a unit cell containing between 2 and 24 magnetic atoms, and (3) the absence of disorder or partially occupied sites. The results are shown in Table~\ref{tab:all_list}.

\begin{longtable*}{ccccccccccc}
\caption{List of materials considered in this paper. Remarks for Formula, $G_{\rm SG}$, and $T_N$ [K] are the same as in Table~\ref{tab:fm-afm1}}\\
\label{tab:all_list}\\
        \toprule[.3mm]
        Formula & $G_{\rm SG}$ & $T_{\rm N}$ [K] &  & Formula & $G_{\rm SG}$ & $T_{\rm N}$ [K] &  & Formula & $G_{\rm SG}$ & $T_{\rm N}$ [K] \\
        \midrule[.15mm]
        \endfirsthead
        
        \multicolumn{11}{c}{\tablename\ \thetable\ (\textit{cont.})} \\\\
        Formula & $G_{\rm SG}$ & $T_{\rm N}$ [K] &  & Formula & $G_{\rm SG}$ & $T_{\rm N}$ [K] &  & Formula & $G_{\rm SG}$ & $T_{\rm N}$ [K] \\
        \midrule[.15mm]
        \endhead

        \endfoot
        \endlastfoot
        
        CsFeO$_{2}$ ht & $Fd\bar{3}m$ & 1055 &  & TbFe$_{6}$Ge$_{6}$ & $Cmcm$ & 489 &  & AuMn$_{2}$ rt & $I4/mmm$ & 366 \\
        RbFeO$_{2}$ ht & $Fd\bar{3}m$ & 1027 &  & TmMn$_{2}$Ge$_{2}$ & $I4/mmm$ & 487 &  & LuMn$_{6}$Sn$_{4}$Ge$_{2}$ & $P6/mmm$ & 366 \\
        KFeO$_{2}$ rt & $Pbca$ & 1001 &  & ScFe$_{6}$Ge$_{6}$ & $P6/mmm$ & 485 &  & FeF$_{3}$ rt & $R\bar{3}c$ & 364 \\
        Fe$_{2}$O$_{3}$ hem & $R\bar{3}c$ & 957 &  & ThMn$_{2}$Si$_{2}$ & $I4/mmm$ & 483 &  & ErMn$_{6}$Sn$_{4}$Ge$_{2}$ & $P6/mmm$ & 362 \\
        Mn$_{3}$N$_{2}$ ht$_{1}$ & $I4/mmm$ & 919 &  & TmMn$_{6}$Ge$_{6}$ & $P6/mmm$ & 482 &  & CaMnSi & $P4/nmm$ & 360 \\
        Dy$_{4}$Ni$_{6}$Al$_{23}$ & $C2/m$ & 900 &  & SrFe$_{12}$O$_{19}$ & $P6_3/mmc$ & 480 &  & TmMn$_{6}$Sn$_{4}$Ge$_{2}$ & $P6/mmm$ & 358 \\
        TiO$_{2}$ ana & $I4_1/amd$ & 880 &  & YbFe$_{6}$Ge$_{6}$ & $P6/mmm$ & 480 &  & Sr$_{2}$FeO$_{3}$F & $P4/nmm$ & 358 \\
        Mn$_{3}$Ir rt & $Pm\bar{3}m$ & 855 &  & YbMn$_{6}$Ge$_{6}$ & $P6/mmm$ & 480 &  & Fe$_{2}$As & $P4/nmm$ & 353 \\
        Mn$_{4}$N rt & $Pm\bar{3}m$ & 750 &  & YMn$_{6}$Ge$_{6}$ & $P6/mmm$ & 476 &  & Au$_{5}$Mn$_{2}$ rt & $C2/m$ & 351 \\
        LaFeO$_{3}$ rt & $Pnma$ & 746 &  & ErMn$_{6}$Ge$_{6}$ & $P6/mmm$ & 475 &  & ErMn$_{6}$Sn$_{6}$ & $P6/mmm$ & 351 \\
        LaFeO$_{3}$ ht$_{1}$ & $R\bar{3}c$ & 740 &  & DyCo$_{3}$Ga$_{2}$ & $P6/mmm$ & 475 &  & ErMn$_{2}$Si$_{2}$ & $I4/mmm$ & 350 \\
        PrFeO$_{3}$ orth & $Pnma$ & 730 &  & TmFe$_{6}$Ge$_{6}$ & $Immm$ & 474 &  & BaY$_{2}$CoO$_{5}$ & $Pnma$ & 350 \\
        La$_{2}$BiO$_{2}$ & $I4/mmm$ & 723 &  & YMn$_{2}$Si$_{2}$ & $I4/mmm$ & 470 &  & GdMn$_{2}$Ge$_{2}$ & $I4/mmm$ & 350 \\
        Ca$_{2}$Fe$_{2}$O$_{5}$ rt & $Pnma$ & 723 &  & LuFe$_{6}$Ge$_{6}$ & $P6/mmm$ & 469 &  & Mn$_{3}$SnC & $Pm\bar{3}m$ & 348 \\
        CrSb & $P6_3/mmc$ & 711 &  & LaMn$_{2}$Si$_{2}$ & $I4/mmm$ & 468 &  & NdMnAsO & $P4/nmm$ & 347 \\
        CaFeO$_{3}$ rt & $Pnma$ & 710 &  & HoMn$_{6}$Ge$_{6}$ & $P6/mmm$ & 466 &  & VO$_{2}$ rt & $P2_1/c$ & 345 \\
        Ba$_{2}$YFe$_{3}$O$_{8}$ tet$_{1}$ & $P4/mmm$ & 690 &  & LuMn$_{2}$Si$_{2}$ & $I4/mmm$ & 464 &  & Fe[NCN] & $P6_3/mmc$ & 345 \\
        EuFeO$_{3}$ & $Pnma$ & 689 &  & MnIrSi & $Pnma$ & 460 &  & FeGe ht$_{2}$ & $C2/m$ & 342 \\
        SmFeO$_{3}$ & $Pnma$ & 688 &  & BaMnGe & $P4/nmm$ & 460 &  & GdMn$_{2}$Si$_{2}$ & $I4/mmm$ & 340 \\
        Sr$_{2}$Fe$_{2}$O$_{5}$ rt & $Ima2$ & 680 &  & FeSe hp$_{1}$ & $P6_3/mmc$ & 460 &  & SmMn$_{2}$Ge$_{2}$ & $I4/mmm$ & 340 \\
        Fe$_{3}$O$_{4}$ rt & $Fd\bar{3}m$ & 679 &  & HfFe$_{6}$Ge$_{6}$ & $P6/mmm$ & 457 &  & YMn$_{6}$Sn$_{6}$ & $P6/mmm$ & 339 \\
        CaMn$_{2}$Ge$_{2}$ & $I4/mmm$ & 670 &  & TbCo$_{4}$B & $P6/mmm$ & 455 &  & Ba$_{2}$MoFeO$_{6}$ ht & $Fm\bar{3}m$ & 338 \\
        KFe[MoO$_{4}$]$_2$ hp$_{3}$ & $P\bar{3}c1$ & 653 &  & Mn$_{3}$Pd$_{5}$ rt & $Cmmm$ & 450 &  & MnNiGe rt & $Pnma$ & 336 \\
        HfMn$_{6}$Ge$_{6}$ & $P6/mmm$ & 650 &  & BaCuYFeO$_{5}$ & $P4mm$ & 448 &  & YMn$_{6}$Sn$_{4}$Ge$_{2}$ & $P6/mmm$ & 336 \\
        FeBiO$_{3}$ hp V & $Pnma$ & 643 &  & Fe$_{7}$Se$_{8}$ ht & $P3_121$ & 447 &  & Ba$_{2}$MoFeO$_{6}$ rt & $I4/mmm$ & 334 \\
        YFeO$_{3}$ rt & $Pnma$ & 639 &  & YbMn$_{2}$Ge$_{2}$ & $I4/mmm$ & 434 &  & NiS rt & $R3m$ & 330 \\
        HoFeO$_{3}$ & $Pnma$ & 637 &  & TbMn$_{2}$Ge$_{2}$ & $I4/mmm$ & 434 &  & LaCrO$_{3}$ ht$_{1}$ & $R\bar{3}c$ & 327 \\
        ErFeO$_{3}$ & $Pnma$ & 636 &  & YMn$_{2}$Ge$_{2}$ & $I4/mmm$ & 431 &  & MnSn$_{2}$ & $I4/mcm$ & 324 \\
        NaFe[MoO$_{4}$]$_2$ & $C2/c$ & 635 &  & DyMn$_{2}$Ge$_{2}$ & $I4/mmm$ & 431 &  & Nd$_{2}$NiO$_{4}$ rt & $Cmce$ & 323 \\
        ErFe$_{4}$B & $P6/mmm$ & 630 &  & BaRuO$_{3}$ rt & $R\bar{3}m$ & 430 &  & Sr$_{2}$MgReO$_{6}$ rt & $I4/mmm$ & 320 \\
        TmFeO$_{3}$ orth & $Pnma$ & 630 &  & TbMn$_{6}$Sn$_{4}$Ge$_{2}$ & $P6/mmm$ & 428 &  & MnAs rt & $P6_3/mmc$ & 318 \\
        YbFeO$_{3}$ orth & $Pnma$ & 629 &  & MnNiO$_{3}$ & $R\bar{3}$ & 427 &  & LaMnAsO & $P4/nmm$ & 317 \\
        FeBiO$_{3}$ rt & $R3c$ & 627 &  & SmMn$_{2}$Si$_{2}$ & $I4/mmm$ & 425 &  & Cr & $Im\bar{3}m$ & 312 \\
        GdCo$_{3}$ & $R\bar{3}m$ & 615 &  & TbMn$_{6}$Sn$_{6}$ & $P6/mmm$ & 422 &  & FeGe rt & $P2_13$ & 310 \\
        TmFe$_{4}$B & $P6/mmm$ & 610 &  & DyMn$_{6}$Ge$_{6}$ & $P6/mmm$ & 421 &  & KFeSe$_{2}$ & $C2/c$ & 310 \\
        NdFeO$_{3}$ & $Pnma$ & 609 &  & CuEu$_{2}$O$_{4}$ rt & $I4/mmm$ & 421 &  & BaCoS$_{2}$ rt mon & $Pmmn$ & 310 \\
        FeS ht$_{2}$ & $P6_3/mmc$ & 599 &  & NdMn$_{2}$Ge$_{2}$ & $I4/mmm$ & 420 &  & TbMn$_{5}$Ge$_{3}$ & $Pnma$ & 310 \\
        FeS tro & $P\bar{6}2c$ & 593 &  & La$_{2}$NiO$_{4}$ lt & $P4_2/ncm$ & 420 &  & LaMnSi & $P4/nmm$ & 310 \\
        HoCr$_{2}$Si$_{2}$ & $I4/mmm$ & 590 &  & DyCo$_{4}$B & $P6/mmm$ & 420 &  & MnTe rt & $P6_3/mmc$ & 309 \\
        FeS ht$_{1}$ & $P6_3mc$ & 588 &  & CaMnGe & $P4/nmm$ & 420 &  & NdCo$_{2}$P$_{2}$ & $I4/mmm$ & 309 \\
        Mn$_{2}$As & $P4/nmm$ & 582 &  & TbMn$_{6}$Ge$_{6}$ & $P6/mmm$ & 419 &  & ScMn$_{6}$Ge$_{6}$ & $P6/mmm$ & 308 \\
        ZrMn$_{6}$Sn$_{6}$ & $P6/mmm$ & 580 &  & CoRh$_{2}$S$_{4}$ & $Fd\bar{3}m$ & 418 &  & CrAs rt & $Pnma$ & 306 \\
        CuFeS$_{2}$ hp & $I\bar{4}$ & 575 &  & PrMn$_{2}$Ge$_{2}$ & $I4/mmm$ & 418 &  & MnRu$_{2}$Ge & $Fm\bar{3}m$ & 305 \\
        HfMn$_{6}$Sn$_{6}$ & $P6/mmm$ & 575 &  & Sr$_{2}$MoFeO$_{6}$ rt & $I4/m$ & 416 &  & Cr$_{2}$O$_{3}$ & $R\bar{3}c$ & 305 \\
        NbFe$_{6}$Ge$_{6}$ & $P6/mmm$ & 561 &  & GdTiGe ht & $P4/nmm$ & 412 &  & PrCo$_{2}$P$_{2}$ & $I4/mmm$ & 304 \\
        TlFeO$_{3}$ & $Pnma$ & 560 &  & Mn$_{3}$Sn & $P6_3/mmc$ & 411 &  & NiS ht & $P6_3/mmc$ & 304 \\
        NdMnSiH & $P4/nmm$ & 560 &  & DyMn$_{6}$Sn$_{4}$Ge$_{2}$ & $P6/mmm$ & 411 &  & BaCuLuFeO$_{5}$ & $P4mm$ & 303 \\
        ErFe$_{6}$Sn$_{6}$ & $Cmcm$ & 560 &  & TbMnSi rt & $Pnma$ & 410 &  & SmCo$_{2}$P$_{2}$ & $I4/mmm$ & 302 \\
        HoFe$_{6}$Sn$_{6}$ orth & $Immm$ & 559 &  & Ca$_{2}$MnFeO$_{5}$ orth$_{1}$ & $Pnma$ & 407 &  & BaY$_{2}$NiO$_{5}$ & $Immm$ & 300 \\
        DyFe$_{6}$Sn$_{6}$ rt$_{-}$orth$_{1}$ & $Cmcm$ & 559 &  & HoCo$_{4}$B & $P6/mmm$ & 404 &  & La$_{3}$Mn$_{4}$Sn$_{4}$ & $Immm$ & 300 \\
        GdFe$_{6}$Sn$_{6}$ & $Cmcm$ & 554 &  & FeGe ht$_{1}$ & $P6/mmm$ & 404 &  & BaNdMn$_{2}$O$_{6}$ ht & $P4/mmm$ & 300 \\
        CuMn$_{3}$As$_{2}$ & $Pnma$ & 550 &  & GdTiSi orth & $P4/nmm$ & 400 &  & HoMn$_{2}$Ge$_{2}$ & $I4/mmm$ & 300 \\
        Mn ht$_{2}$ stab & $I4/mmm$ & 540 &  & LaMn$_{2}$Ge$_{2}$ & $I4/mmm$ & 398 &  & BaMn$_{3}$O$_{6}$ & $C2/m$ & 295 \\
        DyFeO$_{3}$ & $Pnma$ & 540 &  & ScMn$_{6}$Sn$_{4}$Ge$_{2}$ & $P6/mmm$ & 397 &  & CuTb$_{2}$O$_{4}$ & $I4/mmm$ & 295 \\
        Ca$_{2}$ReFeO$_{6}$ lt & $P2_1/c$ & 540 &  & DyMn$_{6}$Sn$_{6}$ & $P6/mmm$ & 395 &  & LaCrO$_{3}$ rt & $Pnma$ & 291 \\
        Sr$_{2}$Co$_{2}$O$_{5}$ & $Ima2$ & 535 &  & Mn$_{3}$SnN & $Pm\bar{3}m$ & 395 &  & Ba$_{2}$Y$_{2}$Co$_{4}$O$_{11}$ rt & $Pmmm$ & 291 \\
        GdFeO$_{3}$ & $Pnma$ & 530 &  & Mn$_{3}$B$_{4}$ & $Immm$ & 394 &  & MnPtAl & $P6_3/mmc$ & 290 \\
        NiO ht & $Fm\bar{3}m$ & 526 &  & CeMn$_{2}$Ge$_{2}$ & $I4/mmm$ & 393 &  & ErMn$_{2}$Ge$_{2}$ & $I4/mmm$ & 290 \\
        Ba$_{4}$Fe$_{9}$O$_{14}$[OH]$_6$ & $C2/m$ & 520 &  & Cr$_{2}$As rt & $P4/nmm$ & 393 &  & Ba$_{2}$Dy$_{2}$Co$_{4}$O$_{11}$ ht & $Pmmm$ & 290 \\
        YbMn$_{2}$Si$_{2}$ & $I4/mmm$ & 520 &  & HoMn$_{6}$Sn$_{4}$Ge$_{2}$ & $P6/mmm$ & 391 &  & SrMnO$_{3}$ 4H & $P6_3/mmc$ & 290 \\
        LuMn$_{6}$Ge$_{6}$ & $P6/mmm$ & 518 &  & MnRhSi & $Pnma$ & 389 &  & CoO ht & $Fm\bar{3}m$ & 289 \\
        Ho$_{6}$Fe$_{23}$ & $Fm\bar{3}m$ & 515 &  & ScMn$_{6}$Sn$_{6}$ & $P6/mmm$ & 388 &  & CoO rt & $I4/mmm$ & 289 \\
        HoCo$_{3}$Ga$_{2}$ & $P6/mmm$ & 514 &  & ErCo$_{4}$B & $P6/mmm$ & 387 &  & TbTiSi & $P4/nmm$ & 287 \\
        TiFe$_{6}$Ge$_{6}$ & $P6/mmm$ & 513 &  & GdMnGe & $Pnma$ & 382 &  & CeCo$_{4}$B & $P6/mmm$ & 287 \\
        TbMnGe & $Pnma$ & 510 &  & FeSn$_{2}$ rt & $I4/mcm$ & 381 &  & TbTiGe rt & $P4/nmm$ & 287 \\
        TbCo$_{3}$ ht & $R\bar{3}m$ & 509 &  & Sr$_{2}$Mn$_{2}$O$_{5}$ orth$_{1}$ & $Pbam$ & 380 &  & CrN lt & $Pmmn$ & 285 \\
        TbMn$_{2}$Si$_{2}$ & $I4/mmm$ & 509 &  & MnCoSi rt & $Pnma$ & 380 &  & CrN rt & $Fm\bar{3}m$ & 283 \\
        ZrFe$_{6}$Ge$_{6}$ & $P6/mmm$ & 509 &  & Sr$_{3}$Fe$_{2}$Cl$_{2}$O$_{4}$ & $I4/mmm$ & 378 &  & AgMn$_{3}$N & $Pm\bar{3}m$ & 281 \\
        Fe$_{3}$BO$_{6}$ & $Pnma$ & 508 &  & CeMn$_{2}$Si$_{2}$ & $I4/mmm$ & 378 &  & SrMnO$_{3}$ 4H rt & $C222_1$ & 280 \\
        LuMn$_{2}$Ge$_{2}$ & $I4/mmm$ & 508 &  & NdMn$_{2}$Si$_{2}$ & $I4/mmm$ & 377 &  & TmMn$_{6}$Sn$_{6}$ & $P6/mmm$ & 280 \\
        DyMn$_{2}$Si$_{2}$ & $I4/mmm$ & 506 &  & HoMn$_{6}$Sn$_{6}$ & $P6/mmm$ & 376 &  & Fe$_{2}$RuSi & $Fm\bar{3}m$ & 280 \\
        GdCo$_{4}$B & $P6/mmm$ & 505 &  & LuMn$_{6}$Sn$_{6}$ & $P6/mmm$ & 373 &  & TiFe$_{2}$ rt & $P6_3/mmc$ & 278 \\
        TmMn$_{2}$Si$_{2}$ & $I4/mmm$ & 498 &  & V$_{2}$WO$_{6}$ & $P4_2/mnm$ & 370 &  & NiBiO$_{3}$ & $P\bar{1}$ & 278 \\
        DyFe$_{6}$Ge$_{6}$ & $Cmcm$ & 497 &  & PrMn$_{2}$Si$_{2}$ & $I4/mmm$ & 369 &  & CuHo$_{2}$O$_{4}$ hp & $I4/mmm$ & 278 \\
        Fe$_{2}$O$_{3}$ orth & $Pna2_1$ & 495 &  & BaTbMn$_{2}$O$_{6}$ rt & $P2_1/m$ & 367 &  & CrSb$_{2}$ & $Pnnm$ & 275 \\
        Sr$_{3}$Cu$_{2}$Fe$_{2}$Se$_{2}$O$_{5}$ & $I4/mmm$ & 490 &  & FeSn & $P6/mmm$ & 367 &  & YFe$_{2}$Si$_{2}$ & $I4/mmm$ & 275 \\
        \bottomrule[.15mm]
\end{longtable*}

\bibliography{refs}

\end{document}